\documentclass[12pt,nofootinbib]{revtex4-1}
\usepackage{amsmath,amssymb}
\usepackage{slashed}

\newcommand\BZ{\mathbb{Z}}
\newcommand{\BC}{\mathbb {C}}
\newcommand{\sech}{\hbox{sech}}

\def\Tr{\textrm{Tr}}

\newcommand{\beq}{\begin{equation}}
\newcommand{\beqs}{\begin{equation*}}
\newcommand{\eeq}{\end{equation}}
\newcommand{\eeqs}{\end{equation*}}

\newcommand{\CP}{{\mathbb {CP}}}

\newcommand{\Ncal}{\mathcal{N}}

\begin{document}
\setlength{\unitlength}{1mm}
\allowdisplaybreaks


\title{Giant gravitons and the emergence of geometric limits in $\beta$-deformations of ${\cal N}=4$ SYM}

\author{David Berenstein, Eric Dzienkowski }
\affiliation { Department of Physics, University of California at Santa Barbara, CA 93106}

\begin{abstract}
We study a one parameter family of supersymmetric marginal deformations of ${\cal N}=4$ SYM with $U(1)^3$ symmetry, known as $\beta$-deformations, to understand their dual $AdS\times X$ geometry, where $X$ is a large classical geometry in the $g_{YM}^2N\to \infty$ limit. We argue that we can determine whether or not $X$ is geometric by studying the spectrum of open strings between giant gravitons states, as represented by operators in the field theory, as we take $N\to\infty$ in certain double scaling limits. We study the conditions under which these open strings can give rise to a large number of states with energy far below the string scale. The number-theoretic properties of $\beta$ are very important. When $\exp(i\beta)$ is a root of unity, the space $X$ is an orbifold. When $\exp(i\beta)$ close to a root of unity in a double scaling limit sense, $X$ corresponds to a finite deformation of the orbifold. Finally, if $\beta$ is irrational, sporadic light states can be present.
\end{abstract}

\maketitle


\section{Introduction }
\label{S:Introduction}

The gauge/gravity duality conjectures that certain gauge field theories are equivalent to quantum theories of gravity in higher dimensions \cite{Maldacena:1997re}. In some of these theories, at large $N$ and strong gauge coupling, the vacuum can be characterized by a classical, geometric supergravity background. In this classical background curvatures are small, gravitational excitations can be treated semi-classically, and non-gravitational states are very massive. It is natural to ask what is the set of such gauge field theories that in the large $N$ limit correspond to classical geometries. 

This question is too hard to answer in general. It has been argued that a parametrically large gap in anomalous dimensions for operators is enough to guarantee the appearance of a geometric dual \cite{Heemskerk:2009pn}, but finding examples where the gap can be followed from perturbation theory to strong coupling is hard. 

In this paper we examine this question of emergent geometry in a very special set of cases. We consider the family of $\beta$-deformations of $\Ncal=4$ SYM theory. This is a continuous family of supersymmetric conformal field theories characterized by two parameters, $g_{YM}^2N$ and $\beta$, which is in turn a subset of the Leigh-Strassler deformations of $\Ncal=4$ SYM \cite{Leigh:1995ep}. In some contexts these are known as $q$-deformations where $q=\exp(2i\beta)$ and $\beta$ is a real number. The classical weakly coupled string theory on a very large geometry arises as the effective description of the undeformed $\Ncal=4$ SYM when $g_{YM}^2 N\to \infty$, and $g_{YM}^2\ll1$. Our main focus is to understand which values of $\beta$ lead to a well defined classical geometry in this limit.

It is known that if $q$ is fixed with $q^s=1$ a primitive $s-$ root of unity, then the above limit corresponds to a $\BZ_s\times \BZ_s$ orbifold with discrete torsion of $AdS_5\times S^5$ \cite{Douglas:1998xa, Berenstein:2000hy}. The orbifold group only acts on the $S^5$, and the geometry depends only on $s$. If we think of  these as $q= \exp(2\pi i t/s)$, for $s,~t$ integers, the volume of the quotient geometry depends only on $s$, but not on $t$.
Such behavior for a physical quantity on the unit circle, characterizing the possible values of $q$, show that the emergence of different geometries is discontinuous in the $\beta$ parameter. After all, the rationals are dense on the unit circle. We will show in this paper that if $\beta$ approaches a root of unity in a particular double scaling limit relative to $g_{YM}^2N$, then we can get a classical geometry which is not just $AdS_5\times S^5/\BZ_s\times \BZ_s$, but rather a supergravity deformation of it.

Notice that the different sphere quotients have very different geometries than a sphere and they are also different from each other in the spectral sense; the spectrum of modes computed on them by restricting states on the $S^5$ to have the appropriate periodicities give different answers. To go from one geometry to another one,  it is necessary to make a discontinuous jump, even though the two dual field theories are connected by a continuous parameter. In string theory, the spectrum of strings should be continuous as a function of $\beta$,  at fixed $g_{YM}^2N\gg 1$, but that does not mean that it is still continuous at $g_{YM}^2N=\infty$.  For finite values of $g_{YM}^2N$, we think  of the system as a stringy geometry, since the sphere is of finite volume in string units in this case \cite{Maldacena:1997re}. The different possible string theory geometries are connected to each other by T-duality \cite{Berenstein:2000ux}, but only one such geometry can become classical at a given time. That is, only one such geometry can become infinite in size in all directions when we take $g_{YM}^2N\to \infty$. This limit can be considered as a boundary of parameter space. What we are saying is that this geometric boundary limit is essentially fractal.
The purpose of this paper is to find a very precise field theory diagnostic of this phenomenon for the particular theories we are studying. 

Naively, just like when one studies T-duality for torus geometries, one should be able to understand the appearance of a large volume geometric limit in terms of a spectrum of closed strings that are much lighter than the string scale, which are the Kaluza-Klein excitations in the given geometry. This spectrum can be characterized by a gap in energies between the Kaluza-Klein scale and the string scale, or equivalently, by having an infrared scale that is very different from the string scale.

The spectrum of closed strings in {\it classical} string theory on $AdS_5\times S^5$ is believed to be controlled by an integrable system \cite{Minahan:2002ve, Bena:2003wd,Dolan:2004ps} and thus in principle is solvable by a Bethe ansatz (see \cite{Beisert:2010jr} for a recent review of the integrability program). 
For the case of $\beta$ deformations, we know that the string theory is integrable; the integrable system arising in field theory results as a twist of the $AdS_5\times S^5$ integrable system \cite{Berenstein:2004ys,Lunin:2005jy,Beisert:2005if} (see also \cite{Zoubos:2010kh} for a review).  As such, tools related to integrability can in principle give us a recipe for the emergence of geometry.  
Unfortunately, this does not automatically translate into a large set of states that are easily followed; the corresponding twisted algebraic equations that need to be solved to find these states are hard to solve for analytically, see however \cite{Frolov:2005ty}. 

For the purposes of this paper we will assume that the integrability program has solved the problem of understanding the geometric origin of $AdS_5\times S^5$ strings, but our goal will be to understand the other geometries that can appear at different values of $q$.

An alternative route to understand the origin of geometry for $T$-duality setups is by using D-branes \cite{Dai:1989ua}. D-branes are geometric objects sometimes described as submanifolds of a given geometry where strings are allowed to end. They can be extended in some directions, and sharply localized in others (in the sense that they can have a well defined position that is sensitive to shorter length scales than the string scale). If an extended D-brane becomes infinitely large in volume as we take the large volume limit in the closed string theory, a collection of open strings ending on the D-brane also become light (they are Kaluza-Klein excitations of gauge fields, etc.) and a similar gap between light states and stringy states characterizes the emergence of geometry. We can also have pairs of parallel D-branes  and study the spectrum of strings stretching between them, testing if they can be considered to be close to each other, or far apart from each other. Indeed, the properties of the  spectrum of such open strings can be used to infer a definition of distance. So if we have such D-branes and can follow the spectrum of such open strings, we can check both that a transverse direction to the D-brane is becoming large, and that a volume of the D-brane is becoming large.

For practical purposes, D-branes are usually better than strings at describing local information. Closed strings usually describe local information by having an S-matrix whose low energy limit can be eventually converted to a local lagrangian. With D-branes, they are already localized in some directions, so local geometric questions are easier to engineer.

This is the route we will take in this paper. There is a set of supersymmetric D-branes in $AdS_5\times S^5$, namely half BPS giant gravitons \cite{McGreevy:2000cw}, which remain BPS states for all $\beta$. 
Once we have chosen which half of the supersymmetries to preserve, these giant gravitons are characterized by coordinates which live on a disk: an angular position, and a radius.
The dual states for these localized giant gravitons were understood recently \cite{Berenstein:2013md} and computations of the spectrum of strings stretching between two such localized states up to two loop order were performed in \cite{Berenstein:2013eya}. An all-loop conjecture was also suggested in this work, where it was argued that these open string states are BPS states with respect to the central charge extension  of the $\mathfrak{su}(2|2)$ spin chain \cite{Beisert:2005tm,Beisert:2006qh}. Because these states are BPS, we will assume there exist dual strings with the same energy in the sigma model. Basically, we are addressing the gap problem \cite{Heemskerk:2009pn} for open strings rather than closed strings.

In this paper we will do three main calculations. First, we will show that the spectrum of open strings stretching between giant gravitons in $AdS_5\times S^5$ can be obtained by truncating classical solutions of the infinite open string constructed in \cite{Spradlin:2006wk}. This provides direct evidence at strong coupling that the open string states can be constructed and that they are indeed BPS.

In the second calculation we will examine precisely how the twist affects the boundary conditions for the spin chain induced by these D-branes. Indeed, any twist can be undone locally by a field redefinition, but the boundary conditions are sensitive to such a redefinition. This means that if we understand the details of the modified boundary conditions, we will still have BPS states relative to the central charge extension of the spin chain. Consequently, we will derive a rich spectrum of states with known energies.

In the third set of calculations we will explore this spectrum of states for various configurations and under which conditions an appropriate double scaling limit exists that still leads to a large geometry, including when $q$ is not exactly a root of unity. 


\section{Field theory results for giant gravitons in $\Ncal=4$ SYM}

Giant gravitons were originally discussed in \cite{McGreevy:2000cw}. These states are spherical D3-branes wrapping an $S^3\subset S^5$ with fixed volume and rotate with angular momentum in the $S^5$. These states preserve half of the supersymmetries and an $SO(4)\subset SO(6)$ of the R-charge. They are mapped to chiral primary scalar operators, implying that they do not carry angular momentum in $AdS_5$. These states are parametrized by a radius and an angle. In a parametrization of the $S^5$ as the collection of unit vectors $\vec x$
\begin{equation}
x_1^2+x_2^2+\dots+x_6^2=1
\end{equation}
we can choose to write the $x_{3,4,5,6}$ coordinates in  a polar decomposition, with a radius and a three sphere. This is the three sphere we choose to wrap the $S^3$ on. The $x_1,x_2$ coordinates belong to a disk, because the radius of the $S^3$ is bounded below by zero. On this disk, the half BPS giant gravitons move with constant angular velocity equal to one (this is always measured relative to the scale of the $AdS$ geometry) and they stay at fixed radius.

Half BPS states are constructed solely from the lowest component of a single chiral scalar field of $\mathcal{N}=4$ SYM. We will label the lowest components of our chiral scalars by $X,~Y,~Z$ and construct our BPS states from $X$ only. We think of each component as corresponding to a complex combination of consecutive pairs of coordinates on the $S^5$: $X\sim x_1 + ix_2$, etc. This identifies the $SO(6)$ R-charge as rotations of the $S^5$.

The dual half BPS states were understood first in \cite{Balasubramanian:2001nh}, where it was argued that they were sub determinant operators built of $X$. Afterwards, it was understood that any gauge invariant operator built out of $X$ is half-BPS to all orders and that a complete orthogonal basis of these states can be obtained from Schur polynomials \cite{Corley:2001zk}, equivalently described by Young tableaux and free fermions. This lets us generalize the set of states to have more than one giant graviton. The next observation made was that these states can also be described in terms of droplets of free fermions in a two-dimensional quantum hall system \cite{Berenstein:2004kk}. Giant gravitons that wrap the sphere are hole states in the fermion droplet. In \cite{Lin:2004nb} it was shown that general fermion droplets can be mapped to supergravity solutions.

To add strings that stretch between giant gravitons we have to go beyond operators built only out of $X$. A general recipe to attach strings to giants in the dual field theory was described in \cite{Balasubramanian:2004nb}. Strings are attached by adding boxes with labels to a half BPS configuration made of Young tableaux. Techniques to deal with basic field theory computations in these states were developed in \cite{de Mello Koch:2007uu,de Mello Koch:2007uv,Bekker:2007ea}.

The states described by Young tableaux have fixed angular momentum. In a quantum system, this means that they are delocalized in the dual angular variable.
To define objects that are localized in the dual angular variable, one needs a superposition of angular momentum states. These geometrically localized states play the role of a coherent background for field theory computations in setups that stretch strings between branes.
Such a formalism was developed in \cite{Berenstein:2013md,Berenstein:2013eya},
where the spectrum of strings was described in terms of an open chain with boundary conditions and was found to be exactly computable.

The basic idea in \cite{Balasubramanian:2004nb} is that an open string is represented as a product of matrices $W = YX^{n_1} \dots Y$, with no trace taken, thought of as a word in the letters $X$ and $Y$.
This word is then attached to a configuration of giants.
We are restricting to the $SU(2)$ sector spanned by $X$ and $Y$. Attaching various such words to a single giant graviton does lead to an approximate Fock space of open strings and a $1/N$ expansion \cite{Berenstein:2003ah}. It is also expected for the more general case where the states in the Fock space are subject to a Gauss' Law. The counting of these restricted string states was produced in a simplified setting in \cite{deMelloKoch:2012ck}.

The usual description of the $SU(2)$ sector is in terms of a ferromagnetic $\mathfrak{su}(2)$ spin chain \cite{Minahan:2002ve} where we take $X\simeq |\downarrow\rangle$ and $Y\simeq |\uparrow\rangle$ at each location of the chain. Because the $X$ letters can jump in and out of the word when we consider the dynamics of the giant graviton, it is better to describe the states of the Hilbert space in terms of an occupation number for $X$ letters between the different $Y$. Thus, we label the states as
\begin{equation}
|n_1, n_2, \dots n_{k}\rangle= Y X^{n_1}Y X^{n_2} \dots X^{n_k} Y
\end{equation}
so they are described by a `boson' at each lattice site. For $k$ lattice sites, there are $k+1$ $Y$ letters in the word $W$. In this setup, the one loop Hamiltonian for the open spin chains can be described in terms of raising and lowering operators for a spin chain of Cuntz oscillators at each site \cite{Berenstein:2005fa}. These are described by the deformed oscillator algebra
\begin{equation}
a a^\dagger =1
\end{equation}
and $a|0\rangle=0$. For general giant gravitons, the one and two loop boundary spin chain Hamiltonian was computed in detail in \cite{Berenstein:2013eya}. The one loop Hamiltonian in the $\mathfrak{su}(2)$ sector for the spin chain Hamiltonian is
\begin{align}
H_{1-\text{loop}} &= \frac{g_{YM}^2 N}{8 \pi^2} \left[ \left( \frac\lambda{\sqrt N}- a_1^\dagger\right)  \left( \frac{\lambda^*}{\sqrt N}- a_1\right)
+(a_1^\dagger - a_2^\dagger) (a_1 -a_2) + \dots \right. \nonumber \\
&\hspace{8cm} \left. + \left( \frac{\tilde \lambda}{\sqrt N}- a_k^\dagger\right)  \left( \frac{\tilde\lambda^*}{\sqrt N}- a_k\right)\right]
\label{eq:1-l}
\end{align}
where $\lambda,~\tilde{\lambda}$ are the complex collective coordinates of the two giant gravitons $\lambda\propto x_1-ix_2$.  They live on a disk of radius $\sqrt N$ \cite{Berenstein:2013md}, and by an $N$ dependent rescaling can be put into a disk of radius one. The normalized coordinates are called $\xi,~\tilde \xi$ respectively, and they end up being rescaled complex conjugates of  the $\lambda$. The conventions on factors of $\pi$ are the same as those in  \cite{Beisert:2003tq}.

The ground state of the $\mathfrak{su}(2)$ one loop open spin chain with $k$ sites is a tensor product of Cuntz oscillator coherent states, one for each site. Each coherent state is defined by $a |z\rangle= z|z\rangle$ with $z$ a coordinate in the complex unit disk. The domain of $z$ is restricted because the states are not normalizable for $|z|\geq 1$. The ground state of the spin chain is characterized by the following equations \cite{Berenstein:2013eya}
\begin{equation}
\label{eq:findif}
z_{i-1} -z_i = z_i- z_{i+1}
\end{equation}
where $z_0 = \xi = \lambda^*/\sqrt N$ and $z_{k+1}= \tilde \xi= \tilde \lambda^*/\sqrt N$. This linear system of equations is completely solvable with
\begin{equation}
z_m = \frac{(k+1-m) \xi+m \tilde \xi}{k+1}
\end{equation}
That is, the sites are linearly interpolated between the two giants.


There is a subtlety when computing the ground state energy. The giant gravitons are built from an order $N$ number of $X$. When we take $N\to\infty$, with $k$ finite, this background contributes infinite energy.
It is standard to remove this infinity by taking the effective Hamiltonian to be $\Delta -J$, where $\Delta$ is the operator dimension, and $J$ is the half-BPS R-charge \cite{BMN}. At this order, we have that
\begin{align}
\Delta-J &= (k + 1) + \langle H_{1-loop}\rangle + \langle H_{2-loop}\rangle+\dots \\
\label{eq:twoloop}
&= (k + 1) + \frac{g_{YM}^2 N}{8\pi^2} \frac{|\xi-\tilde\xi|^2}{(k+1)} + \frac{(g_{YM}^2N)^2}{128\pi^4}\frac{|\xi - \tilde{\xi}|^4}{(k+1)^3} + \dots
\end{align}

In \cite{Berenstein:2013eya} it was argued that the all-loop energy of such a string resulted from a square root expression
\begin{equation}
\Delta -J = \sqrt {(k+1)^2 +\frac{g_{YM}^2 N}{4 \pi^2}| \xi-\tilde \xi|^2}\label{eq:allloop}
\end{equation}
consistent with their result \eqref{eq:twoloop}.
In the field theory this energy is the dimension of an operator. On the gravity side, this is an energy measured in $AdS$ radius units. Their argument was that this open string appears to be a BPS state for the centrally extended $\mathfrak{su}(2|2)$ symmetry of the (infinite) spin chain discovered in \cite{Beisert:2005tm,Beisert:2006qh}. The central charge is exactly given by 
\begin{equation}
\Xi = \xi - \tilde{\xi}
\end{equation}
and specifies the energy of the correpsonding state. This central extension is characterized by adding or removing one $X$ to a given infinite spin chain, first on the left and  then on the right; basically it sends $Y\to [X,Y]$. Acting on an impurity at fixed quasi-momentum $p$ on a ferromagnetic ground state (keeping all other impurities fixed in position), the result is proportional to $\exp(i p)-1$. For closed strings the central charge is confined. However, just like in flat space, the central charge extension of a supersymmetry algebra can be sourced by D-branes, so that the open string states stretching between them are short massive multiplets. The shortening of representations is standard for parallel branes in flat space. 

One may also try to understand the central charge construction from a Bethe ansatz perspective.
As discussed by Dorey in \cite{Dorey:2006dq}, bound states in the $\mathfrak{su}(2)$ spin chain are BPS.
This is derived from a Bethe ansatz calculation. What we want to do is check that our ground state has a similar local form to a Bethe ansatz.

To do this, we first need to expand the coherent state ground states in terms of the occupation number basis
\begin{equation}
|z\rangle= N_z \sum z^n |n\rangle
\end{equation}
where $N_z$ is a normalization factor. Now, when we look at the spin chain ground state in the interior (away from the boundaries) the $X$ spins can jump to the left or to the right of $Y$. To jump an $X$ from the right of the $n$-th $Y$ defect to the left, we need to take $|s\rangle_{n}\rightarrow |s-1\rangle_{n}$ and at the same time, we need to take $|t\rangle_{n-1}\rightarrow |t+1\rangle_{n-1}$. The relative cost in amplitude for these two components of the wave function is $z_n/z_{n-1}\simeq \exp(i p_n)$, which can be interpreted as the momentum of the $n$-th $Y$ defect in the background of the $X$, so each $Y$ has a unique well defined momentum in the coherent state. In a Bethe ansatz state, we would expect that such ordering of momenta can only happen if the relative momenta are at a zero for the $S$-matrix between defects.
The scattering matrix of two $Y$ with momenta $k_1,~k_2$ has a pole at $\exp(-i k_1)+\exp(ik_2)=2$ (see \cite{Berenstein:2004ys} for conventions and details). This translates to the equations
\begin{equation}
\frac{z_{n-1}}{z_n}+\frac{z_{n+1}}{z_n}=2
\end{equation} 
which is equivalent to 
\begin{equation}
z_{n-1}+z_{n+1}-2z_n=0
\end{equation}
which is exactly the  equation \eqref{eq:findif}. This shows that the local structure of the one loop wave function is that of a bound state of magnons in a Bethe ansatz sense. This should persist to all orders.

Since these bound states have been identified with classical solutions of the string sigma model \cite{Spradlin:2006wk}, we can conjecture that the corresponding open string solutions for the open strings are locally the same, but where the classical solutions are cut at the location of the giant gravitons. We will show this in the next section.


\section{Open strings between giant gravitons}

In this section we want to compute the energy of a string stretching between two giant gravitons in $AdS_5\times S^5$ with fixed angular momentum on the $S^5$ to confirm the conjecture proposed in \cite{Berenstein:2013eya}. To do so, we need to solve the string sigma model on $AdS_5\times S^5$ with boundary conditions given by the giant graviton. Since we expect these strings to be BPS (they are expected to be short representations of supersymmetry), they should correspond to solutions of the sigma model that have this property. 

The BPS states in $\mathcal{N}=4$ SYM with large angular momentum in the infinite chain limit are bound states of elementary excitations as shown by Dorey \cite{Dorey:2006dq} called giant magnons. Each state is characterized by the quantum numbers
\begin{align}
\Delta - J_1 &= \sqrt{Q^2+\frac\lambda{\pi^2} \sin^2(p/2)} \\
J_2&= Q
\end{align}
with $Q$ the number of magnon excitations, $J_i$ the Cartan generators of the $SO(6)$ R-charge, and $p$ the quasi momentum of the bound state.
In particular, $J_1$ is the R-charge of the half-BPS ground state which is infinite in the infinite chain limit. The corresponding classical string solutions were found in \cite{Spradlin:2006wk,Kalousios:2006xy} (see also \cite{Chen:2006gea}). There, the $J_i$ correspond to the angular momenta of the string on the $S^5$ and $p$ is the geometric angle subtended by the string stretched between two points on the edge of a maximal disk in the $S^5$. Like their spin chain dual, these strings have infinite length. These dual BPS strings satisfy a simpler set of equations that the full set of states. Since our open spin chains are also BPS, their dual string states should also satisfy a simple set of equations, albeit modified to include the effects of the new boundary conditions. Thus we take our ansatz to be these classical string solutions but truncated so that the endpoints fall on the giant gravitons.

We verify that these truncated solutions are dual to the open spin chains in three parts. First we need to relate the positions of the giants gravitons in the $S^5$ to the collective coordinates of the giant gravitons in the open spin chain. Otherwise we could not compare the energy and angular momentum of the string to that of the open spin chain. Then we need to check that the infinite string can be truncated so that the ends follow the positions of the giant gravitons (this is, the string ends don't fall from the giant graviton). Lastly, we need to relate $\Delta - J_1$ and $J_2$ for the truncated solutions and get the answer conjectured in \cite{Berenstein:2013eya}, namely
\begin{align}
\Delta -J_1 &= \sqrt{Q^2+\frac\lambda{4 \pi^2}|\xi - \tilde \xi|^2}  \\
J_2&= Q
\end{align}
where $\xi,~\tilde \xi$ are coordinates on a unit disc.

First we need to find a relation between $\xi,~\tilde{\xi}$ and the positions of the giant gravitons.
Let us label the coordinates on the sphere by $z_i$, $i=1,2,3$ with the constraint $\sum_i \bar{z}_iz = 1$. The corresponding angular momenta are the $J_i$. If we have a giant graviton with angular momentum $L$, wrapping an $S^3$ inside $S^5$ (it is a point in the $z_1$ coordinate), then it moves with angular velocity equal to one and sits at $|z_1| = \sqrt{1-L/N}$ \cite{McGreevy:2000cw}. In the dual CFT, in terms of the description based on fermion droplets and coherent states of the dual field theory, these also sit on a disk which can be normalized to have radius one \cite{Berenstein:2004kk,Berenstein:2013md}, and have energy 
\begin{equation}
E = L = N(1 - |\xi|^2)
\end{equation}
Fixing $L$, we see that $|\xi| = |z_1|$. Furthermore, both notions of the angle direction between field theory and gravity are the same; the angle is related to the conjugate variable to angular momentum.

In order to impose the truncated boundary conditions, we need to make sure that the string solutions can end on the giant gravitons and that they rotate with angular speed equal to one in the $z_1$ plane. This follows because the ends of the infinite string classical solution travel at the speed of light and moreover they reside at the edge of the disk. Traveling at the speed of light in this case means that they move with angular velocity one and if we cut the solutions, they actually end on the giant gravitons.

Using the variable definitions of \cite{Spradlin:2006wk}, we recall the classical infinite string solution. The worldsheet coordinates are called $t,~x$. The time on the world sheet is identified with the external time variable. The solution depends on two auxiliary variables $\alpha,~\theta$. They are related to the magnon excitation number $Q$ and quasi momentum $p$ by
\begin{align}
\cot(\alpha) &= \frac{2r}{1 - r^2}\sin\left(\frac{p}{2}\right) \\
\tan(\theta) &= \frac{2r}{1 + r^2}\cos\left(\frac{p}{2}\right)
\end{align}
where $r$ is a function of the energy, angular momenta, quasi momentum, and boundary conditions.
The classical solution for the infinite string is then given by
\begin{align}
u &= (x\cosh(\theta) - t\sinh(\theta))\cos(\alpha) \\
v &= (t\cosh(\theta) - x\sinh(\theta))\sin(\alpha) \\
z_1 &= e^{it}\left(\cos\left(\frac{p}{2}\right) + i\sin\left(\frac{p}{2}\right)\tanh(u)\right)
\label{eq:z1} \\
z_2 &= e^{iv}\frac{\sin\left(\frac{p}{2}\right)}{\cosh(u)}
\end{align}
where $u,~v$ are auxiliary functions.
Notice that at $t=0$ the real part of $z_1$ is constant.

As we can see from equation \eqref{eq:z1}, the string moves at constant angular velocity equal to one. In these solutions, the range of $u$ is infinite. For our case, we want to cut the string so that the range of $x$ (or $u$ for that matter) is finite.

To find our string, we need to fix the boundary conditions by restricting the range of the variables so that $z_1(u_1,t)=\xi \exp(it)$ and $z_1(u_0,t)= \tilde \xi \exp(it)$ and then solving for $u_0,~u_1$. Since we need the real part of $z_1$ constant at $t=0$,  we choose $\xi,~\tilde \xi$ to have the same real part, which can be achieved by a global rotation of the D-brane configuration. The boundary conditions now read
\begin{align}
\cos\left(\frac{p}{2}\right) + i\sin\left(\frac{p}{2}\right)\tanh(u_1) &= \xi  \\
\cos\left(\frac{p}{2}\right) + i\sin\left(\frac{p}{2}\right)\tanh(u_0) &= \tilde{\xi} 
\end{align}
The quasi momentum $p$ is found from the real part of $\xi$ or $\tilde \xi$.
Consequently we have
\begin{equation}
\tanh(u_1) = \frac{\xi - \cos(p/2)}{i\sin(p/2)}, \quad \tanh(u_0) = \frac{\tilde{\xi} - \cos(p/2)}{i\sin(p/2)}
\end{equation}
so that $\xi,~\tilde{\xi}$ are enough to determine $p$ and the values of $u_0,~u_1$.
We also have the relation
\begin{equation}
{\sin\left(\frac{p}{2}\right)}(\tanh(u_1) - \tanh(u_0)) = {|\xi - \tilde{\xi}|}
\end{equation}
which will be imporant when computing $\Delta - J_1$.

The energy and the angular momentum are given by 
\begin{align}
\Delta - J_1 &= \frac{\sqrt{\lambda}}{2\pi} \int dx\, (1 - \Im(\bar{Z}_1\partial_t Z_1)) \\
J_2 &= \frac{\sqrt{\lambda}}{2\pi}\int dx\, \Im(\bar{Z}_2\partial_t Z_2)
\end{align}
over the appropriate range. This is evaluated at $t=0$ for convenience, where $u,~x$ are proportional to each other.

After some lengthy algebraic computation, one shows that
\begin{align}
\Delta - J &= \frac{\sqrt{\lambda}}{2\pi}\frac{1 + r^2}{2r}\sin\left(\frac{p}{2}\right)\cosh(\theta)\cos(\alpha)\int dx\, \sech^2(u) \\
&= \frac{\sqrt{\lambda}}{2\pi}\frac{1 + r^2}{2r}\sin\left(\frac{p}{2}\right)\int du\, \sech^2(u) \\
&= \frac{\sqrt{\lambda}}{2\pi}\frac{1 + r^2}{2r}\sin\left(\frac{p}{2}\right)(\tanh(u_1) - \tanh(u_0))
\end{align}
Similarly, one finds that the $J_2$ angular momentum is given by
\begin{align}
J_2 &= \frac{\sqrt{\lambda}}{2\pi}\sin\left(\frac{p}{2}\right)\cosh(\theta)\cos(\alpha)\int dx\, \sech^2(u) \\
&= \frac{\sqrt{\lambda}}{2\pi}\frac{1 - r^2}{2r}\sin\left(\frac{p}{2}\right)\int  du\, \sech^2(u) \\
&= \frac{\sqrt{\lambda}}{2\pi}\frac{1 - r^2}{2r}\sin\left(\frac{p}{2}\right)(\tanh(u_1) - \tanh(u_0))
\end{align}
We now want to fix $J_2=Q$. Since we know $p,~u_0,~u_1$,  this determines the value of $r$. 

Notice that apart from the $r$ dependence, $\Delta-J_1$ and $J_2$ are essentially identical. 
Call $A_{\pm}= \frac{1 \pm r^2}{2r}$. One then sees that $A_+^2-A_-^2=1$. It follows that
\begin{align}
(\Delta -J)^2 - J_2^2 &= \left (\frac{\sqrt{\lambda}}{2\pi}\right )^2 \left|\sin\left(\frac{p}{2}\right) (\tanh u_1-\tanh u_0)\right|^2(A_+^2-A_-^2)\\
&= \left (\frac{\sqrt{\lambda}}{2\pi}\right )^2 \left|\sin\left(\frac{p}{2}\right) (\tanh u_1-\tanh u_0)\right|^2\\
&= \left (\frac{\lambda}{4\pi^2}\right ) |\xi-\tilde \xi|^2
\end{align}
So we find that 
\begin{equation}
\Delta - J_1 = \sqrt{Q^2 + \frac\lambda{4 \pi^2}|\xi-\tilde \xi|^2}
\end{equation}
as expected.


\section{Marginal deformations of $\Ncal=4$ SYM, protected operators and the $SU(2)$ sector}

The ${\cal N}=4  $ SYM admits a three parameter (complex) family of deformations that preserves ${\cal N}=1$ super  conformal invariance \cite{Leigh:1995ep}. If we use $\Ncal=1$ SUSY notation, the $\Ncal=4$ SYM has three chiral fields $\tilde X,~\tilde Y,~\tilde Z$ in the adjoint  representation of the gauge group. We will be interested in the case where the gauge group is $SU(N)$, and then $\tilde X,~\tilde Y,~\tilde Z$ are $N\times N$ matrices of superfields.
We will reserve the letters $X,~Y,~Z$ without tildes for the  lowest component of the corresponding superfields, in the following manner:
\begin{equation}
\tilde X(x, \theta) = X(x)+ \sqrt 2\theta \psi_X(x) +\theta^2 F_X(x)
\end{equation}
The three parameter family is characterized by the superpotential
\begin{equation}
W= \lambda_1\Tr[\tilde X,\tilde Y] \tilde Z+\lambda_2\Tr (\{\tilde X,\tilde Y\} \tilde Z)+ \lambda_3\Tr(\tilde X^3+\tilde Y^3+\tilde Z^3) 
\end{equation}
where $\lambda_1,~\lambda_2,~\lambda_3$ are arbitrary complex parameters. 
The gauge coupling constant is determined from $\lambda_1,~\lambda_2,~\lambda_3$ by requiring that the beta function of the gauge coupling constant vanish. This amounts to the vanishing of the anomalous dimension of the fields $X,~Y,~Z$. Because the superpotential has enough discrete symmetries to cyclically replace $\tilde X\to \tilde Y\to\tilde Z\to \tilde X$, the anomalous dimensions of $\tilde X,~\tilde Y,~\tilde Z$ are identical and they need to be set to zero. This requires the $R$-charge of $\tilde X,~\tilde Y,~\tilde Z$ equal to one.
 In that case the beta function vanishes. This requirement of vanishing anomalous dimension is enough to determine $g_{YM}$ in terms of $\lambda_1,~\lambda_2,~\lambda_3$. When $\lambda_3=0$, the superpotential has a $U(1)^3$ symmetry, where $\tilde X,~\tilde Y,~\tilde Z$ can be independently rotated from each other, a linear combination of these rotations, combined with spinor rotations is the $U(1)_R$ charge. The $\Ncal=4$ theory is obtained when $\lambda_2=\lambda_3=0$, and in that case $\lambda_1=g_{YM}$ (this convention requires the $X,~Y,~Z$ fields to be canonically normalized). We are interested in the case where we turn on both $\lambda_1$ and $\lambda_2$. It is convenient to rewrite 
the superpotential as follows
\begin{equation}
W= G_{YM}  \left[ \Tr(\tilde X\tilde Y\tilde Z) - h\Tr(\tilde X\tilde Z\tilde Y)\right]
\end{equation} 
 where $G_{YM}$ is a function of $g_{YM}$, $h$. When $h^*h=1$, we call $h=\exp(2 i\beta)$, and in this case $G_{YM}=g_{YM}$. For the particular case of roots of unity, these result from orbifolds with discrete torsion of the $\Ncal=4$ SYM theory itself \cite{Douglas:1998xa,Berenstein:2000hy}. In that case, planar diagrams match those of the ${\cal N}=4 $ SYM. This can be conveniently understood by noticing that the corresponding spin chain can be obtained by twisting the original spin chain of ${\cal N}=4$ SYM theory \cite{Berenstein:2004ys}. Results to various loop orders when $|h|\neq 1$ were obtained in \cite{Elmetti:2007up}. At one loop order one has that $g_{YM}^2 (1+|h|^2) = 2 G_{YM}^2$. This is easy to obtain by requiring that the one loop planar anomalous dimension of $X$ be equal to zero. The study of general $h$, although rather interesting, is beyond the scope of the present paper. This is because, as argued in \cite{Berenstein:2004ys}, the model fails to be integrable. For us this means that we can not control the results of calculations to all orders in perturbation theory.
 
In this paper we will be mostly interested in the generalized $SU(2)$ sector. This is spanned by gauge invariant operators made of $X,~Y$ alone \footnote{$X,~Y$ are the lowest components of the superfields  $\tilde X,~\tilde Y$}, which for $h=1$ corresponds to a sector of operators that is closed under perturbation theory and that moreover has an $SU(2)$ symmetry of rotations of $X$ into $Y$. This $SU(2)$ is part of the $SU(4)$ R-charge of ${\cal N}=4 $ SYM. For general $h$, operators with these quantum numbers can only mix with each other as there are no other states with the same quantum numbers (this necessarily includes the bare dimension of the operator).

Let us identify the $U(1)^3$ symmetry charges of the theory. In our conventions the $\theta,~\bar\theta$ variables of superspace have $U(1)_R$ charge equal to $\mp 3/2$ respectively, and $\tilde X$ has $R$-charge equal to one. Thus, a superpotential term in the action of the form 
\begin{equation}
\int d^2\theta \Tr(\tilde X^3)
\end{equation}
has net R-charge equal to zero. Also, for the vector superfields we use $W_\alpha\simeq \psi_\alpha +\theta(F+D)_\alpha+\theta^2(\bar{\slashed{\partial}}\bar \psi)_\alpha$. 
We can classify our fields by the $U(1)^3$ charges. This is depicted in table \ref{tab:charges}. 
For the conjugate fields we reverse all the charges, except the dimension $\Delta$. We also introduce the charge $J$ in analogy with the corresponding charge for ${\cal N}=4 $ SYM \cite{BMN}. Fields other than $X$ in the field theory have $\Delta -J>0$.
\begin{table}[b]
\begin{tabular}{|c|c|c|c|c|c|}
\hline
Field&$U(1)_R$& $U(1)_1$& $U(1)_2$&$\Delta$& $(\Delta-J)$\\ \hline
$X$& $1$ & $1/2$&0&1&0\\
$Y$& $1$ & $-1/2$& $1/2$&1& 1\\
$Z$& $1$ & $0$& $-1/2$&1&1\\
$\psi$ & $3/2$ & $0$ & $0$&3/2&1\\
$\theta$ & $-3/2 $& $0$ &$0$&-1/2&0\\
\hline
\end{tabular}
\caption{List of charges of the fields under the $U(1)^3$ symmetry and the dimensions of the fields. For other fields they are determined by the charge of the superspace variables and requiring that superfields have definite charges.
We have also defined $J= (1/3) Q_R+ (4/3) Q_1+(2/3) Q_2$.  
}\label{tab:charges}
\end{table}

A particular subclass of operators is the set that are made of the single field $X$. Any polynomial of $X$ that is gauge invariant is identical in algebraic form to those of the half-BPS sector of ${\cal N}=4$ SYM that would be built out of $X$. These are multitraces of $X$. If the degree of the polynomial is $n$, then the R-charge of such an operator is $n$, and the $U(1)_1$ charge is $n/2$.  All of these operators have $\Delta-J=0$. Thus, this subsector does not mix with any other (all other sectors have fields with $\Delta-J>0$). We also have that these operators can not have any anomalous dimension either. After all, these operators are the lowest component of a scalar chiral superfield $f(\tilde X)$, so $\bar D  f(\tilde X)=0$. These equations imply that either the lowest component of the multiplet $f(\tilde X)$ is a primary field, invariant under half of the supersymmetries, and therefore its dimension is entirely controlled by the $U(1)_R$ charge, or that $f(\tilde X)$ is a descendant
of the form $
f(\tilde X) =\bar D U$ for some other superfield $U$ that is not chiral. Remember that $\bar D$  acts without changing the value of $\Delta-J$. The right hand side would need to be an operator with $\Delta-J=0$, of the same class as the ones we are studying. But these operators are chiral, so no such $U$ superfield can give a non-zero right hand side. Only the first possibility is available: the operator has a protected dimension. This is independent of how many traces make up the operator. This is also true for arbitrary values of $h$, not only those where $|h|=1$. This means that this entire sector is protected by supersymmetry, no matter the trace structure of the operator.

The set of half BPS operators can be classified in the free field theory limit by studying Young tableaux \cite{Corley:2001zk}, and then the arguments in \cite{Berenstein:2004kk} show that the collection of these objects can also be visualized in terms of a quantum hall droplet (free fermions on a magnetic field). We can also interpret the corresponding particle  and hole states as analogous to dual giant gravitons and giant gravitons. Configurations can have one such giant graviton or many. Also, the collective coordinate formalism of \cite{Berenstein:2013md} can be adapted to this problem without any modifications. Hole states will be associated to a point inside a fermion droplet of radius $\sqrt N$ \footnote{This is the radius of the droplet for a particular normalization of the collective coordinate, it can also be rescaled to be of order $1$}.

As described before, we are interested in the analogue of the $SU(2)$ sector. These are operators built of $X,~Y$.  The operator $\Delta-J$ will count the number of $Y$ fields, which we will call $m$. Also, the $U(1)_2$ charge of this operator is equal to $ m/2$. There are not many other operators that carry those quantum numbers for $\Delta-J$ and $U(1)_2$. The only candidate that also carries positive values of $U(1)_2$ is $\bar \psi_Z$. Such a field contributes also one to $\Delta -J$ and $1/2$ to $U(1)_2$. However, it has a smaller $R$ charge than $Y$, namely, its R-charge is equal to $1/2$, rather than one, so it can  not mix with the operators we have described (all three conserved quantum numbers would need to match for mixing to occur). This means that the $SU(2)$ sector survives as a sector for all values of $h$, and not just for $h=1$.

We will first consider single trace operators in the asymptotic limit where we take
\begin{equation}
{\cal O}\simeq \sum_{[n_i]} a^{[n_i]}\Tr(X^{n_1} Y X^{n_2} \dots X^{n_k} Y)
\end{equation}
in the planar limit with $k$ fixed and very high occupation numbers $n_i$. We are interested in the value of $\Delta - J$ for the operator $\cal O$ and when it exists. This requires that $\sqrt N\gg\sum n_i\gg k$. We can take $\sum n_i \sim N^{1/a}$ with $a>3$ for example to ensure that we can stick to planar diagrams. In this limit the $Y$ defects are dilute, and we can treat them independently of each other on a first approximation. The spin chain has translation invariance of the $Y$ relative to the position of the $X$, essentially because of the cyclic property of the trace \cite{BMN}. The vacuum $\Tr(X^n)$ is a ferromagnetic ground state with minimal $\Delta-J$. A single $Y$ defect is an impurity. In the free field theory limit $G^2_{YM}N\to 0$, $Y$ contributes to $\Delta-J$ by one. When there is more than one $Y$, the $n_i$ labels matter and serve to measure the distance between the defects. At large separation we have approximate translation invariance. Thus, it is natural that asymptotically we have that  $a^{[n_i]}\simeq \exp( i (q_1 n_1+q_2 n_2\dots +q_k n_k))$,  so that each defect carries some quasi momentum equal to $\phi_i = q_{i+1}-q_i$.  In the planar limit, at each given loop order $s$, the $Y$ can move at most $s$ steps to the left or to the right, so the energy (this is the same as the anomalous dimension) of a defect can be computed locally and should depend only on $\phi_i= q_{i+1}-q_{i}$. 
We expect therefore to have a dispersion relation $\Delta -J = \sum E(\phi)$, where each defect contributes $E(\phi)$ to the energy.

An important question for us is what is the form of $E(\phi)$ for arbitrary $h=\exp(2i\beta)$ and $g_{YM}$. This has been answered for $h=1$ in a variety of ways \cite{Santambrogio:2002sb,Berenstein:2005jq,Beisert:2006qh}. The main result is that
\begin{equation}
E(\phi) = \sqrt{1 +\frac {g_{YM}^2 N}{\pi^2} \sin^2(\phi/2)} 
\end{equation}
In the notation of \cite{Santambrogio:2002sb}, the quantity $4 \sin^2(\phi/2)= -\alpha= \exp(i\phi)+\exp(-i\phi)+2= -(1-\exp(i\phi))(1-\exp(-i\phi))$ is determined by the equations of motion. 
The same calculation in the presence of $\beta$ leads to
\begin{equation}
E(\phi) = \sqrt{1 +\frac {g_{YM}^2 N}{\pi^2} \sin^2(\phi/2-\beta)} 
\end{equation}
so that one effectively shifts the quasi momentum of each particle excitation in the spin chain  from $\phi\to \phi-2\beta$. This is in accordance with the fact that the associated spin chain is just a twist of the original one \cite{Berenstein:2004ys}. One should also notice that this superpotential can be written in terms of a generalized star product \cite{Lunin:2005jy},  so one can guarantee that the planar diagrams of the $\beta$-deformed theory coincide with the planar diagrams of ${\cal N}=4 $ SYM, to all orders, up to the point where we care about the periodicity conditions of the various fields on the spin chain. A calculation for those giant magnons  based on the sigma model can be found in \cite{Bobev:2006fg, Bykov:2008bj}.
 For general $|h|\neq 1$, there is no integrability at one loop level \cite{Berenstein:2004ys}, and one can not  argue that a deformed $\mathfrak{su}(2|2)$ symmetry survives that would protect the result on the right hand side. 

Our goal in this paper will be to understand the corresponding energy of the $\mathfrak{su}(2)$ ground state with $n$ copies of $Y$ and arbitrary $X$, for an open string whose ends attach to a giant graviton made of $X$. This energy is interpreted as a dispersion relation for a fluctuation between the D-branes with $n$ units of momenta.


\section{The $\beta$-deformed spin chain with boundaries}

Following  the calculations of \cite{Berenstein:2004ys} it is easy to write down the closed spin chain Hamiltonian for the one loop $SU(2)$ sector in the Cuntz oscillator basis for arbitrary $h$. 
The answer is
\begin{align}
H_{1-loop}&=\frac{G_{YM}^2 N}{8 \pi^2}\sum_n 
(a_n^\dagger - h a_{n+1}^\dagger) (a_n - h^* a_{n+1})\label{eq:hham}\\
&= \frac{g_{YM}^2 N}{8 \pi^2}\sum_n 
(\exp (-i\beta) a_n^\dagger - \exp(i \beta) a_{n+1}^\dagger) (\exp(i\beta)a_n -\exp(-i\beta) a_{n+1}) \label{eq:hhambeta}
\end{align}
the second line is specific to $h=\exp(2i\beta)$.
This is directly derived from the superpotential. Roughly, the cost to switch $XY\to YX$ in the equations of motion of $Z$ is a factor of $h$. At this loop order, only the square of the superpotential shows up. In this form the Hamiltonian is also a sum of squares, and in the $SU(2)$ sector for arbitrary $h$ it corresponds to the XXZ chain.

The Hamiltonian \eqref{eq:hhambeta} is obtained by replacing ordinary commutators in the dilatation operator by $q$-deformed commutators.
That is
\begin{equation}
\Tr([X,Y][\partial_Y,\partial_X]) \rightarrow \Tr((XY- q YX)(\partial_Y\partial_X - q^* \partial_X\partial_Y))
\end{equation}
Because we are working in the $\beta$-deformed theory with gauge group $SU(N)$ and not $U(N)$, there is an additional term that needs to be added.
\begin{equation}
\label{eq:qdeformeddil}
\Tr([X,Y][\partial_Y,\partial_X]) \rightarrow \Tr((XY - q YX)(\partial_Y\partial_X - q^* \partial_X\partial_Y)) - \Tr(XY - q YX)\Tr(\partial_Y\partial_X - q^* \partial_X\partial_Y)
\end{equation}
The added term yields an additional contribution to the anomalous dimension in the planar limit only for the operator $\Tr(XY)$, in which case the anomalous dimension vanishes entirely.
In \cite{Fokken:2013mza} it was argued that the only state affected in the $SU(2)$ sector by these finite size effects, called prewrapping, is $\Tr(XY)$.
Additionally it was shown that this state is protected to all loop orders.
Integrability, which we have assumed captures both sides of the AdS / CFT correspondence in the planar limit, predicts a divergent anomalous dimension for $\Tr(XY)$ \cite{Arutyunov:2010gu}.
Later on we will only be considering operators with large R-charge  and so these finite size corrections unaccounted for by integrability for very short closed strings are not a concern to us.
We do not need to make any modifications to $H_{\text{1-loop}}$.

To add the boundary conditions, we follow the calculations in \cite{Berenstein:2013md,Berenstein:2013eya}.
The way this works with the giant graviton collective coordinates amounts to adding phases to the collective coordinates $\lambda$ for the boundary contributions.
We get that 
\begin{align}
H_{1-loop} &= \frac{g_{YM}^2 N}{8 \pi^2}  \left[ \left( \exp(-i\beta) \frac\lambda{\sqrt N}-\exp(i\beta) a_1^\dagger\right)  \left( \exp(i\beta) \frac{\lambda^*}{\sqrt N}- \exp(-i\beta) a_1\right)+\dots\right. \nonumber
\\ 
&\qquad +\left. \left( \exp(i\beta) \frac{\tilde \lambda}{\sqrt N}-\exp(-i\beta) a_k^\dagger\right)  \left( \exp(-i\beta) \frac{\tilde\lambda^*}{\sqrt N}-\exp(i\beta) a_k\right)\right]
\end{align}
The twist to turn the theory to the previous spin chain is easier to explain in the open spin chain. We just replace $a_s=  \exp(-2i s \beta) \tilde a_s$. Notice that this is an automorphism of the Cuntz algebra, if at the same time we take  $a^\dagger_s=  \exp(2i s \beta) \tilde a^\dagger_s$. We can then easily check that the phases cancel in the spin chain hamiltonian after this replacement. This should be thought of as a local field redefinition of the local fields $a_n,~a_n^\dagger$.
To include the boundary conditions,  we just need to take the result in equation \eqref{eq:1-l} and make the replacement
\begin{equation}
\tilde \lambda\to q^{k+1}\tilde \lambda= \exp[2i (k+1)\beta] \tilde \lambda
\end{equation}

From here, it is easy to find the energy of the open string ground state with angular momentum $n=k+1$ to all orders. We just copy the result in equation \eqref{eq:allloop} with the appropriate substitutions. We find that 
\begin{equation}
\label{eq:q-defdis}
\Delta - J = \sqrt{n^2 + \frac{g_{YM}^2N}{4\pi^2} |\xi - q^{-n} \tilde \xi|^2}
\end{equation}
The power of $q^{-n}=q^{-k-1}$ arises because $\lambda$ and $\xi$ are related to each other by complex conjugation.
This result follows from putting together two observations: one based on integrability of the spin chain, and one based on just field theory arguments. The two observations are that equation \eqref{eq:allloop} is correct (due to the central charge extension symmetry of the all loop spin chain) and  that the field theory dynamics predicts planar equivalence (with a twist) of the ${\cal N}=4$ SYM spin chain for the $\beta$-deformed version. This is the usual statement that noncommutative field theories and regular field theories have the same 
planar diagrams, which in this case results from a $*$-product deformation \cite{Lunin:2005jy}. These statements can be made entirely within quantum field theory and do not require additional insight from string theory.

Consider the operators of the $SU(3)$ sector with $\ell_1$ $Y$ and $\ell_2$ $Z$ defects against an $X$ background. In ${\cal N}=4$ SYM at one loop order, these operators can be obtained by an $SO(4)$ rotation of the ground state with only $Y$ defects. In the twisted theory, the net twist of the boundary condition is proportional to $q^{\ell_1-\ell_2}$. The equations of motion of the Leigh-Strassler theory are cyclic in $X,~Y,~Z$. Thus the cost in phase for a $Z$ to get past an $X$ ends up being opposite in phase to the cost of having a $Y$ jump past an $X$.
 
For this more general case we find that
\begin{equation}
\Delta - J = \sqrt{(\ell_1+\ell_2)^2 + \frac{g_{YM}^2N}{4\pi^2} |\xi - q^{-\ell_1+\ell_2} \tilde \xi|^2}
\label{eq:q-defdis2}
\end{equation}
We can also understand similar $SO(4)$ rotations of $Y$ into $\bar Z$ and $Z$ into $\bar Y$, and the corresponding twists. For our purposes, the difference between $Y,~Z$ is enough.


\section{Geometric limit interpretation}

In this section we consider various space-time configurations of D-branes, choosing suitable values of $\lambda,~\tilde \lambda$ (or $\xi$, $\tilde{\xi}$), and ask what happens to the spectrum in the limit\quad $g^2_{YM}N\to\infty$ using expressions \eqref{eq:q-defdis} and \eqref{eq:q-defdis2}.
The main question we will ask is which states remain light in this limit. We will call states light if their energy is below the typical string scale $\Delta-J<\ell_s^{-1}= (g_{YM}^2 N)^{1/4}$. 
Even though $\Delta,~J\simeq O(N)$ for the giant graviton ground state, it always remains light in this sense since $\Delta - J = 0$. The appearance of $\ell_s$ makes sense because we are measuring energies in units of the $AdS$ radius in the gravity theory. As such, $\ell_s$ can be thought of as a dimensionless ratio of the string length to the $AdS$ radius.

As a warm up to analyze this problem, let us start in the undeformed ${\cal N}=4 $ SYM with a giant graviton at $\xi$, and another one at $\tilde \xi$.
Equivalently we are considering the case where $q = 1$.
The spectrum of states between them will contain light states if
\begin{equation}
\Delta-J=\sqrt{n^2+\frac{g_{YM}^2N}{4\pi^2} |\xi-\tilde\xi|^2} < (g_{YM}^2 N)^{1/4}
\end{equation}
unpacking the inequalities, we need that both
\begin{equation}
n^2 < (g_{YM}^2N)^{1/2}
\end{equation}
and that
\begin{equation}
\frac{g_{YM}^2N}{4\pi^2} |\xi-\tilde\xi|^2<(g_{YM}N)^{1/2}
\end{equation}
because the term in the square root is a sum of squares.

The first term tells us that the momentum of the state is below the string scale, this is $n<\ell_s^{-1}$.  This is usually what we mean by a low energy limit. Notice that this condition is satisfied for all fixed $n$ when we take the limit $g_{YM}^2N\to \infty$. This shows that the volume of the five sphere is becoming infinite in string units; more and more modes are available below the string scale as we take the geometric limit.

The second term tells us essentially  that
\begin{equation}
|\xi-\tilde \xi| < \ell_s
\end{equation}
so that the two D-branes have to be closer to each other than the string scale. This is a standard way to extract the low energy field theory in the Maldacena limit \cite{Maldacena:1997re}. Notice that  in this case all field theory modes survive (there is one state per angular momentum $n$, up to the degeneracy expected from supersymmetry).  Moreover, this second term in the sum of squares can be thought of as the Higgs mass that results when we separate two D-branes by a distance $|\xi-\tilde \xi|$.

The simplest way to understand the geometric location of these states is to consider maximal giant gravitons first, and to consider the standard fibration structure of  the 5-sphere as a circle bundle over  the complex projective plane, $S^1\to S^5 \to \CP^2$. This fibration determines a choice of an ${\cal N}=1 $ superspace $R$-charge.
BPS chiral ring states have their energy equal to the angular momentum along the $S^1$, which means that in the geometric optics limit they are at a fixed position in $\CP^2$ (they are null geodesic in $AdS_5\times S^5$). The little group of such fixed  position determines a fixed $SU(2)\times U(1) \subset SU(3)$ decomposition. The state carries no $SU(2)$ quantum numbers, so that it can be interpreted as a highest weight of $SU(3)$ with respect to this geometric decomposition. This is one way to think of building $\CP^2$ in terms of coherent states.

In the presence of a maximal giant graviton, which is a maximal $S^3$ that shares its $S^1$ fiber with that of the $S^5$, the allowed positions for such an open string lie in a $\CP^1\subset \CP^2$, and similarly can be interpreted as a highest weight of $SU(2)$. Seen from the point of view of the Hopf vibration of $S^3$, 
an object that carries $n$ units of angular momentum on $S^3$ along the Hopf fiber can be thought of as a highest weight state for a monopole spherical harmonic of charge $n$ on the base. These highest weight states are localized on $\CP^1$ because the effective magnetic field is proportional to $n$ and the Landau level classical orbits are circles centered around some position on the sphere. The angular momentum is along the direction of the point on the $\CP^1$.

A similar statement can be made for the other giant gravitons. These end up moving at constant speed on the $S^1$ fiber of $S^5$, and the string is moving along with them. It also moves inside the $S^3$, and the $SU(2)$ chiral symmetry preserved by the giant graviton determines a similar Hopf fibration of this $S^3$. Indeed, it results from looking at the $x_3, \dots x_6$ coordinates of the $S^5$ as a $\BC^2$, and then we choose the standard complex structure on this $\BC^2$ to pick the $SU(2)$ we need.

The next case we want to look at are the orbifolds with discrete torsion, where $q$ is no longer equal to one, but instead is a fixed root of unity; let us say $q^s=1$ is a primitive root of unity for some integer $s > 1$. This implies that $\beta$ is rational. These orbifolds are interesting because the geometry is given by a $S^5/\BZ_s\times \BZ_s$ quotient \cite{Douglas:1998xa,Berenstein:2000hy}. A giant graviton at fixed $\xi$ will correspond to a brane wrapping a $S^3/\BZ_s$ (we think of it as the corresponding $S^3$ in the covering space $S^5$ and act by the corresponding quotient group that maps the position of the brane to itself, otherwise we think of it by the method of images a la Douglas-Moore \cite{Douglas:1996sw} ).  

There are two interesting questions to ask. First, let us ask what happens at generic values of $\tilde \xi=\xi$; we are considering the spectrum of fluctuations of a single brane. To have a light state will then require that 
\begin{equation}
n < \ell_s^{-1}
\end{equation}
and that
\begin{equation}
g_{YM}^2 N |\xi|^2 |1- q^{-n}|^2 <(g_{YM}^2 N)^{1/2}
\end{equation}
For generic $|\xi|\simeq 1$,  we will need that $|1-q^{-n}|\to 0$ , or equivalently, that $n$ is a multiple of $s$; only fluctuations with $s$ units of angular momentum will survive the low energy limit.
More precisely, if we go to the $SU(3)$ sector, we will require that $|1-q^{-\ell_1+\ell_2}|$ survive, which shows that $\ell_1-\ell_2$ is a multiple of $s$. This is exactly what we expect from the optical limit on a $S^3/\BZ_s$ space, which decomposes as a Hopf vibration with a $\CP^1/\BZ_s$ base. The other heavy states can be thought of as long strings stretching between $\xi$ and its images $q^k \xi$.

Similarly, we can ask what happens if we take $\tilde \xi=q^m \xi$, that is, we try to locate a second brane in one of the image points of $\xi$ under the $\BZ_s$ action that can act on the complex coordinate $X$. 
What we find in that case is that a light spectrum of states between the branes survive, so long as $n$ differs from $m$ by a multiple of $s$.
This spectrum of states has fixed differences in momenta, plus a shift from zero. The natural interpretation is that the two D-branes are on top of each other, but they differ in the choice of discrete electric Wilson lines between them, in a similar vein to \cite{Gukov:1998kn}. This means that the coordinate $\xi$ contains both the position and the Wilson line information. The position is uniquely determined by $\xi^s$. Obviously, we can also take limits where $|\tilde \xi - q^{-n}\xi|<\ell_s$ to have such setups, and the interpretation in terms of a relative discrete Wilson line does not change.

A natural question is to ask how we can deal with magnetic Wilson lines, along the lines of \cite{Gukov:1998kn}. This would be important to understand S-duality on the set of states. It is hard to understand the S-dual magnetic strings between branes. See however \cite{Berenstein:2009qd}, where it is argued that the D-strings and $(p,q)$-strings have the same world sheet sigma model as the ordinary strings, except for their tension. We would expect that the BPS central charge argument is extended to these as well, with the tension of the string making an appearance inside the square root formula.  However, a field theory computation for these states is beyond what can be done with perturbation theory. Also, the action of S-duality on the Leigh-Strassler deformations is complicated \cite{Dorey:2002pq}. For us, the D3-branes with magnetic flux need to be at the same location, so they must have the same value of $\xi^s$. Hence they should be linear combinations of the branes at the values of $\xi$ and its images. It is natural to assume that such branes with magnetic Wilson lines  will have fixed values of the R-charge modulo $s$ and are related to the ones with fixed $\xi$ by a  discrete Fourier transform. Understanding S-duality in detail is beyond the scope of the present paper.

Now let us ask also about the special limit where $|\xi|\to 0$. In this case all of these states can survive. In that case, we ask that $|\xi| < \ell_s$ and we get a construction where the images of a brane are separated from a brane itself by distances that are sub-stringy. The spectrum is then the same as that of a single maximal giant graviton. We can think of it as an $S^3$ world volume, or as a $U(s)$ theory on $S^3/\BZ_s$ in a ground state where there is a discrete nonabelian Wilson line given  by
\begin{equation}
W\simeq ( 1, q, q^2, \dots q^{s-1})
\end{equation}
All we need is that $W^s=1$ as a matrix. Both of these give the same spectrum of states. Which is more appropriate will then depend on the nature of the local interactions; if they are local on $S^3$, or on $S^3/\BZ_s$. We will not answer this question here.

Now let us consider $\beta$ to be close to zero, that is $q\to 1$, and $\xi=\tilde \xi$. Again, we are asking about fluctuations of a single brane. In this case we can Taylor expand in $\beta$ around $q=1$.
\begin{equation}
q^n \simeq 1+2i \beta s n + O(n^2)
\end{equation}
Again, just as before, we ask that $n<\ell_s^{-1}$, and that 
\begin{equation}
g_{YM}^2 N |\xi|^2 \beta^2 n^2 <( g_{YM}^2 N)^{1/2}
\end{equation}
One simple way to do this is to take $\beta^2 (g_{YM}^2 N)$ finite. 
It can even be made very large in a double scaling limit sense so long as we allow ourselves to restrict $n$ to be smaller.

What we find is that the energy is proportional to $n$, and more generally, to the square root of a quadratic form involving $\ell_1,~\ell_2$.
This is the dispersion relation of a squashed sphere, where different directions have been squashed differently. Keeping $\beta^2 (g_{YM}^2 N)$ gives a finite squashing; the sphere is still of a size comparable to the $AdS$ radius.

In this case, if we also separate the branes slightly, taking $\xi \neq \tilde \xi$, we notice that the dispersion relation becomes a square root of a quadratic form involving $\ell_1,~\ell_2$ plus a constant term, and more crucially, a linear term will arise. One can even fix $\tilde \xi$ as being related to $\xi$ by a phase such that the factors of $q$ cancel for some $n$.
Such a linear term is like a position dependent relative Wilson line. This will need to be interpreted as having a non-trivial H-flux in the geometry (this follows similar reasoning to  \cite{Berenstein:2000te}). These should end up matching the Lunin-Maldacena geometries when we explore them in more detail \cite{Lunin:2005jy}, where the squashing of the sphere and the $H$ flux is known. 
This problem of reading the flux can also be analyzed using other techniques with D-brane instantons \cite{Ferrari:2013pq}.

The next question we need to ask is what happens for irrational $\beta$. At least naively, nothing survives. This is because the numbers $1-q^{-n}$ will typically never be close enough to zero at finite $n$.  However, if we let $\tilde{\xi}=\xi$ with $|\xi|$ close to zero, we can have states for which $|\xi|^2 |1-q^{-n}|^2<\ell_s^2$. If we use a continuous fraction approximation to $\beta/\pi$, we can find integers $r,~t$ such that $|\beta/\pi - r/t| < 1/ t^2$. We then find that $|\xi|$ can be larger than $\ell_s$ by a factor of roughly $t$. Then the states whose momenta are multiples of $t$ will survive to low energies, so long as the first order Taylor expansion of the exponential in $1/t$ is still a reasonable approximation. Thus, for sufficiently small values of $|\xi|$, even if they are larger than the string scale by a factor of $t$ they look similar to orbifolds with discrete torsion $S^5/\BZ_t\times \BZ_t$. The value of $t$ changes as we go away from the fixed point $|\xi|=0$. In this case we jump between geometric duality frames depending on the distance from the origin $|\xi|=0$. Even if the full result is not geometric, certain classes of questions could be asked in the corresponding orbifold with discrete torsion.

Notice also that if $\ell_1=\ell_2$, the states always survive. This means that we should think of the brane as having at least one large circle of radius one in $AdS$ units, with the other directions forming a stringy geometry. The general structure of states is very similar to what happens in the study of Melvin models \cite{Kutasov:2004aj}, where the different rational approximations to an irrational number play an important role. In our case we are dealing with open strings stretched between D-branes, rather than with the closed string spectrum. It is natural to imagine that the closed string sector in these $\beta$-deformed theories will also have a list of sporadic light states that depend on the number theory properties of $\beta$. A natural difference is that in the work \cite{Kutasov:2004aj} the light states came from wrapped strings on a small circle, while in our case they carry angular momentum. Momentum versus wrapping are T-dual to each other, so exploring these issues further is very interesting.

A natural question to ask is how much of this picture could be obtained from a stringy computation.  Given the Lunin-Maldacena geometries, the corresponding giant graviton states are known \cite{Pirrone:2006iq,Imeroni:2006rb}, even in more general deformations where they become unstable \cite{deMelloKoch:2005jg}. Seeing as our classical solutions stretching between branes correspond to cutting well known solutions of the sigma model on $S^3$, it should be possible to produce these from solutions in the TsT transformations that generate the Lunin-Maldacena backgrounds and the work \cite{Frolov:2005dj}. These should also be applied to the orbifolds with discrete torsion. These details are beyond the scope of the present paper.


\section{Conclusion}

In this paper we have analyzed the spectrum of open strings between giant gravitons in both the ${\cal N}=4 $ SYM theory at strong coupling and in the  $\beta$-deformations of ${\cal N}=4 $ SYM theory. We argued that this set of open string states can be understood as a set of BPS states for the central charge extension of the infinite spin chain discovered by Beisert \cite{Beisert:2005tm} that determines their energy. 

Apart from previous two loop results suggesting this central charge argument \cite{Berenstein:2013eya}, in this paper we also provided classical open string states between giant gravitons with the corresponding energy and quantum numbers. We also argued that these states locally look very similar to those that appear in the Bethe ansatz.

These results can be ported over to the $\beta$-deformed theories because the spin chain is almost the same as in ${\cal N}=4$ SYM, except for a twist which only affects the boundary conditions of the open strings in a simple way based on the quantum numbers of the string states. The energies of these strings are encoded simply in equations \eqref{eq:q-defdis} and \eqref{eq:q-defdis2} which makes it possible to take limits and understand geometry very simply. We argue that physics can be interpreted geometrically in the strong coupling limit if there is a rich set of open string states with low energies (low compared to the string scale) that survives.

This strong coupling limit with a large set of states depends very strongly on the number theoretic properties of $\beta$, and the notion of geometry jumps discontinuously as we move in $\beta$ at infinite coupling, in a way that is very reminiscent of Melvin models \cite{Kutasov:2004aj}. In general, we expect a similar  structure as a function of $\beta$ for all toric quivers, since their dual geometries can also be deformed by the Lunin-Maldacena method \cite{Lunin:2005jy}. It would also be interesting to understand other marginal deformations of ${\cal N}=4 $ SYM and which of them are geometric. A lot less is known about such cases (see however \cite{Berenstein:2002ge, Walton:2011aq}, where examples with dual geometries are expected). 

It would be interesting to explore this issue further in other orbifolds of ${\cal N}=4 $ SYM. This again results in the same spin chain, but the twistings that need to be done are different \cite{Beisert:2005he} and it would be interesting to see how this can affect the study of giant graviton states. It's also interesting to explore $N=2$ theories, where we also expect a central charge extension to control the allowed energies of the states, but where we don't expect integrability \cite{Gadde:2010ku}.

We also discovered that in the orbifolds with discrete torsion, the giant gravitons carried electric Wilson lines on their world volume. This suggests interesting questions regarding how S-duality acts on those states. Considering that the action of S-duality is complicated when looking at different questions \cite{Dorey:2002pq} for $\beta$-deformed theories, this suggests that resolving these issues might be very non-trivial.


\acknowledgments

D.B. would like to thank Greg Moore for various comments.
Work supported in part in part by DOE under grant DE-SC0011702.
E. D. supported  by the Department of Energy Office of Science Graduate Fellowship Program (DOE SCGF), made possible in part by the American Recovery and Reinvestment Act of 2009, administered by ORISE-ORAU under contract no. DE-AC05-06OR23100. D.B. thanks the Aspen Center for Physics for hospitality while completing this work. The ACP is supported in part by the National Science Foundation under Grant No. PHYS-1066293.

\end{document}